\documentclass[12pt, prd, showpacs]{revtex4}
\usepackage{amssymb}
\usepackage{amsmath}
\usepackage[dvips]{graphicx}
\usepackage{epsfig}
\usepackage{color}

\input{tcilatex}

\begin{document}

\title{On particle dynamics near the singularity inside the Schwarzschild \
black hole and T-spheres}
\author{A. Radosz}
\affiliation{Faculty of Basic Problems of Technology (Wroclaw),Wroclaw University of
Science and Technology, 50-370Wroclaw, Poland}
\email{andrzej.radosz@pwr.edu.pl}
\author{A. V. Toporensky}
\affiliation{Sternberg Astronomical Institute, Lomonosov Moscow State University and
Kazan Federal University, Kremlevskaya 18, Kazan 420008, Russia}
\email{atopor@rambler.ru}
\author{O. B. Zaslavskii (corresponding author)}
\affiliation{Department of Physics and Technology, Kharkov V.N. Karazin National
University, 4 Svoboda Square, Kharkov 61022, Ukraine}
\email{zaslav@ukr.net}

\begin{abstract}
The problem of the speed of the objects inside the Schwarzschild black hole
is considered. The general result is that the value of the relative speed of
the objects following their non-zero angular momentum trajectories, both of
geodesic and non-geodesic character, when approaching the ultimate
singularity, tends to the value of speed of light. There is only one
exception when both objects move in the same plane and have parallel angular
momenta. This outcome appears to have a deeper sense: it reflects the
anisotropic character of the dynamics of interior of this particular black
hole. The result in question means that near the singularity, collisions of
two particles lead to an indefinitely large energy in the center of mass
frame. Aforementioned properties have their counterpart in the phenomenon of
an indefinitely large blueshift near the singularity. Thus the angular
momentum of a particle turns out to be an important feature that affects the
final behavior of particle near the singularity. Motivated by this fact, we
generalize the Lema\^{\i}tre frame under the horizon in such a way that
reference particles themselves have nonzero angular momentum. Our results
apply not only to the Schwarzschild singularity but also to other space-like
ones for which the scale factor $g\rightarrow \infty $. We also analyze
another type of singulairites for which the circumference radius vanishes
but $g$ remains finite.
\end{abstract}

\maketitle

\section{Introduction.}

The studies of the properties of the strong gravitational fields and in
particular the properties of the black holes (BH) have got in a recent
decade a significant not only theoretical but also experimental impact. The
first ever picture of the BH namely, supermassive M-87 BH \cite{87},
gravitational waves emission following two BHs merger \cite{merg} and the
intriguing temporarily varying radiation emission of the accretion disk of
the sources of the strong gravitational field \cite{ed} - \cite{jac} are the
most important recent experimental aspects of the presence of the strong
gravitational fields. The theoretical studies of black holes have been also
continuing. Among other interesting things, a special role is played by the
so-called Ba\~{n}ados-Silk-West (BSW) effect: under certain conditions,
two-particle collisions undergoing in the vicinity of the BH's horizon would
lead to the unbounded energy release \cite{ban}.

Meanwhile, there exists another singled out space-time region. This is the
vicinity of the singularity. For the naked singularity it was shown that
high energy collisions are indeed possible there (see, e.g. \cite{kerrnaked}%
). However, in the case of black holes it is hidden beyond the horizon. We
show that a similar effect near the singularity happens even without
rotation, i.e. for the Schwarzschild black hole. Such collisions can, in
principle, change the fate of a black hole and potentially lead to a new
type of the horizon instability due to backreaciton of particles on metric.

For spherically symmetric space-times, there exists also another type of
space-like singularity, apart from the Schwarzschild one. It occurs in the
metric of so-called T-spheres \cite{rub}. The geometry near the singularity
is highly anisotropic both for the Schwarzschild metric and T-sphere but in
the first case one of two scale factors tends to infinity whereas the second
one vanishes. For T-spheres one of two scale factors remains finite.
Collisions of massive particles is the counterpart of red/blue shift inside
a black hole when corresponding particles are massless (photons). It leads
to a bright ring around a singularity (see page B-25 of \cite{TW}). A
careful analytical study of red/blue shift for such massless particles have
been recently done in \cite{gr20}. In doing so, the crucial role of angular
momenta reveals itself that changes overall picture radically as compared to
a pure radial motion (see e.g. \cite{ham}, \cite{gg}).

In recent paper \cite{rad} a general radially freely falling frames are
described and the formula for 3-velocity with respect to these frames have
been derived. If one particle is comoving with the frame in question, the
considered 3-velocity of the second particle is just a relative velocity of
such particles. If we are going further and describe a mutual motion of two
particles, both having non-zero angular momentum, we can proceed in two
ways. First, we can use known formulae for a radially falling frame and
apply them to both particles, after that extract their mutual motion from
this information. In such a way it is reasonable to use the simplest
possible choice for the frame. A natural choice is the static frame. We
should however mention that a velocity with respect to this frame has
particular properties, for example, the radial velocity of a particle
following geodesic and approaching the horizon tends to that of light, $%
V\rightarrow 1$, but after crossing the horizon the speed turns out to be
decreasing to zero \cite{speed}, as if the test particle being hampered
inside horizon.

It is worth noting that this non-monotonic behavior of a 3-velocity is a
specific property of the observers, static outside and resting inside
horizon \cite{speed} and does not appear in the frame connected with Lema%
\^{\i}tre coordinates. Indeed, it was shown that the 3-velocity with respect
to the Lema\^{\i}tre frame at a horizon can take any value from $0$ to $1$,
and this velocity decreases in a monotonic way (if it is not equal to zero
identically) in the Schwarzschild black hole reaching $0$ at a singularity
(if the angular momentum is zero) \cite{Zero}. When we are interested in the
asymptotics near a singularity the behavior of the velocity near horizon
does not matter, so we use the static frame.

As for the second possibility, we can consider a frame connected with the
first particle, thus, falling non-radially. This frame is convenient when we
are interested not only in asymptotic but in the overall particle motion
under the horizon.  This requires generalization of formulae from \cite{rad}
for non-radially falling frames. Thus in the present paper we make a next
step in which we generalize the results of the previous paper and get the
most general formulae for velocity with respect to an arbitrary freely
falling frame.

The paper is organized as follows. In Sec. II we define the line element and
the tetrad for resting observer inside Schwarzschild BH and we apply them
(Sec. III) for description of the kinematics of a freely falling test
particle. In Sec. IV the anisotropy of the this space-time is described.
Particle collisions are analyzed in the following three sections: V -
general setup, VI - in-plane collisions, VII - collisions of particles
moving in different planes. In Secs. VIII-IX the effects of action of
external force are described. In Sec. X we analyze another type of
singularity when the scale factor in the longitudinal direction remains
finite. In Sec. XI we generalize the Lem\^{a}itre frame to include reference
particles with nonzero angular momentum. In Sec. XII we compare properties
of local velocity with respect to a frame with those of non-local velocity,
defined as a time derivative of a proper distance between two different
points. Discussion and final remarks are presented in the final section.

\section{Metric, tetrad}

Let us consider the black hole metric

\begin{equation}
ds^{2}=-fdt^{2}+\frac{dr^{2}}{f}+r^{2}d\omega ^{2}.  \label{1}
\end{equation}%
Here, $f(r_{+})=0$, where $r_{+}$ is the radius of the event horizon. Our
main concern is the Schwarzschild metric for which $f=1-\frac{r_{+}}{r}$,
though we need not to fix $f$ for most of our results. Inside the horizon,
the mutual role of temporal, $t$ and spatial, $r$, coordinates interchanges.
We can choose $T=-r$, $y=t$, where $-r_{+}\leq T\leq 0$, $-\infty <y<\infty $
\cite{nov61}. Then 
\begin{equation}
ds^{2}=-\frac{dT^{2}}{g}+gdy^{2}+T^{2}d\omega ^{2}\text{,}  \label{met}
\end{equation}%
collisions of particles moving in different planes where $g=-f$.

In what follows, it is convenient to use the tetrad attached to a resting
observer with constant spatial, $y,\theta ,\phi $ coordinates. Such an
observer follows a geodesic that has no analogue in the outer part of
space-time \cite{dor}. Namely, in the coordinates ($T,y,\theta ,\phi $) 
\begin{equation}
\hat{h}_{(0)\mu }=-\frac{1}{\sqrt{g}}(1,0,0,0),  \label{0}
\end{equation}%
\begin{equation}
\hat{h}_{(1)\mu }=(0,\sqrt{g},0,0),
\end{equation}%
\begin{equation*}
\hat{h}_{(2)\mu }=(0,0,\left\vert T\right\vert ,0)
\end{equation*}

\begin{equation}
\hat{h}_{(3)\mu }=(0,0,0,\left\vert T\right\vert \sin \theta ).  \label{3}
\end{equation}

\section{Motion of a free particle}

Outside the horizon there exists the time-like Killing vector that
corresponds to the time translation and it leads to the conservation of a
particle's energy. Inside the horizon it becomes a space-like one and this
leads to the conservation of the $y$-component of momentum. The angular
momentum is conserved everywhere.

Then, the four-velocity $u^{\mu }$ of a particle that moves within the plane 
$\theta =\frac{\pi }{2}$ has the form in the coordinate system (\ref{met})%
\begin{equation}
u^{\mu }=(P\text{,}-\frac{p}{g},0,\frac{L}{T^{2}})\text{,}  \label{u}
\end{equation}%
where%
\begin{equation}
P=\sqrt{p^{2}+g(\frac{L^{2}}{T^{2}}+1)},  \label{momentum}
\end{equation}%
$p=-u_{y}$ is the specific conserved momentum (the sign minus is chosen to
keep the maximum similarity with the region outside the horizon), $L=u_{\phi
}$ is the specific, conserved angular momentum.

The corresponding tetrad components read 
\begin{equation}
u^{(a)}=u^{\mu }h_{\mu }^{(a)}=(\frac{P}{\sqrt{g}},-\frac{p}{\sqrt{g}},0,%
\frac{L}{\left\vert T\right\vert })\text{.}
\end{equation}

Then, one can obtain (see Sec. 8 in \cite{flows}) that the three-velocity
has components%
\begin{equation}
V^{(1)}=-\frac{p}{P}\text{,}  \label{v1}
\end{equation}%
\begin{equation}
V^{(3)}=\frac{L\sqrt{g}}{\left\vert T\right\vert P}.  \label{v3}
\end{equation}%
The absolute value of the velocity $V=\sqrt{\left( V^{(1)}\right)
^{2}+\left( V^{(3)}\right) ^{2}}$,%
\begin{equation}
V=\sqrt{1-\frac{g}{P^{2}}}.
\end{equation}

The Lorentz gamma factor%
\begin{equation}
\gamma =\frac{1}{\sqrt{1-V^{2}}}=\frac{P}{\sqrt{g}}.  \label{ga1}
\end{equation}

For given $p$ and $L$, the velocity under discussion obeys the condition%
\begin{equation}
\frac{V^{2}}{1-V^{2}}=\frac{p^{2}}{g}+\frac{L^{2}}{T^{2}}.  \label{V}
\end{equation}

If the singularity is being approached, $T\rightarrow 0$, $g\rightarrow
\infty $.

Then, if $L\neq 0$, we have%
\begin{equation}
\left\vert V\right\vert \approx 1-\frac{1}{2}\left( \frac{T}{L}\right) ^{2}.
\end{equation}

In doing so, $V^{(1)}\rightarrow 0$, $V^{(3)}\rightarrow \pm 1$.

If $L=0$, $V^{(3)}=0$. In the limit under discussion $V^{(1)}\rightarrow 0$.

Thus, in any case $V^{(1)}\rightarrow 0$.

\section{Geometry and dynamics}

The above result can be \ given the following geometric interpretation. The
space-time described by the line element (2) may be referred to as a $T$%
-"sphere" \cite{rub}. It has got some particular properties: it is
non-static, homogeneous, finite in time extent. It has a hypercylindrical a
space-like section $V^{3}=R^{1}\times S^{2}$ with no symmetry center, open ($%
-\infty <y<\infty $)\ in radial, homogeneity direction $R^{1}$. This may be
regarded as an anisotropic cosmological model, expanding longitudinally and
contracting transversely in a two-sphere $S^{2}$ of radius $\left\vert
T\right\vert $ (see also \cite{dor} - \cite{rad19}). Expansion along $y$
axis is finally getting extremely violent: all of the objects are carried
away in such a manner that their "own" speeds are getting negligible - they
are finally in a relative rest. And that is the meaning of the first of the
results of the former section: the resting (in $y$ axis) observer measures
the speed of the test object travelling along this axis as diminishing to
zero, $V\rightarrow 0$, as is seen from (\ref{V}) when $L=0$ and the
singularity is approached, so $g\rightarrow \infty $.

If the velocity vector of a test object has got a transverse (to $y$ axis)
component, i.e. its angular momentum is non-zero, $L\neq 0$, it is also
carried away transversely due to the transverse contraction. This transverse
contraction is of critical character: hypercylinder $V^{3}$ collapses to the
line as the radius of the two-sphere tends to zero. All of the massive and
massless particles are carried away in the following manner. The value of
the speed of the massive test objects as measured by resting observers goes
to that of light, $V\rightarrow 1,$ and the light rays (massless test
objects) are perceived by the resting observer as indefinitely blueshifted 
\cite{gr20}.

Then, the following interesting question arises: what is relative speed of
the two observers depending on their angular momenta? If particles collide,
whether their energy in the center of mass frame remains finite or grows
indefinitely? In particular, it concerns particles travelling with (a)
parallel, (b) antiparallel angular momenta. These questions are considered
below.

\section{Particle collisions: general setup}

Now, we consider collisions of two particles of masses $m_{1}$ and $m_{2}$
and \ briefly analyze the behavior of the energy $E_{c.m.}$ in the center of
mass at the point of collision. By definition,%
\begin{equation}
E_{c.m.}^{2}=-P_{\mu }P^{\mu },
\end{equation}%
where $P^{\mu }=m_{1}u_{1}^{\mu }+m_{2}u_{2}^{\mu }$ is the total
four-momentum. Then,%
\begin{equation}
E_{c.m.}^{2}=m_{1}^{2}+m_{2}^{2}+2m_{1}m_{2}\gamma _{12},
\end{equation}%
where $w$ has the meaning of the relative speed, the Lorentz factor of
relative motion%
\begin{equation}
\gamma _{12}=-u_{1\mu }u^{2\mu }=\frac{1}{\sqrt{1-w^{2}}}  \label{ga}
\end{equation}%
should not be confused with the individual gamma factor of each particle (%
\ref{ga1}).

Below, we discuss two cases separately.

\section{Particles move in the same plane}

Then, it follows from (\ref{u}) and (\ref{ga}) that

\begin{equation}
\gamma _{12}=\frac{P_{1}P_{2}-p_{1}p_{2}}{g}-\frac{L_{1}L_{2}}{T^{2}},
\label{ga12}
\end{equation}

It is instructive to describe collisions in terms of kinematic
characteristics. One can define the angle $\psi $ between particles 1 and 2
according to%
\begin{equation}
\cos \psi =\frac{\vec{V}_{1}\vec{V}_{2}}{V_{1}V_{2}}\text{,}
\end{equation}%
where $\vec{V}_{1}\vec{V}_{2}=V_{1}^{(1)}V_{2}^{(1)}+V_{1}^{(3)}V_{2}^{(3)}$%
. Then, it follows from (\ref{v1}), (\ref{v3}) that%
\begin{equation}
\cos \psi =\frac{1}{\sqrt{p_{1}^{2}+g\frac{L_{1}^{2}}{T^{2}}}\sqrt{%
p_{2}^{2}+g\frac{L_{2}^{2}}{T^{2}}}}(p_{1}p_{2}+\frac{L_{1}L_{2}g}{T^{2}})%
\text{,}
\end{equation}%
\begin{equation}
\gamma _{12}=\gamma _{1}\gamma _{2}(1-\cos \psi )\text{.}  \label{gacos}
\end{equation}

Our main concern is the behavior of $\gamma _{12}$ near the singularity. For
fixed $\,L_{1},$ $L_{2}$, the absolute velocity of each particle in the
limit when the singularity is approached, can take only two values: either $%
V=0$ or $V=1$ \cite{rad}, \cite{flows}. Below, we enumerate different
sub-cases separately depending on the angular momentum of each particle.

\subsection{ $L_{1}=0=L_{2}$.}

Then,%
\begin{equation}
P=\sqrt{p^{2}+g}.
\end{equation}%
For $g\rightarrow \infty $ we have%
\begin{equation}
\gamma _{12}\approx 1+\frac{w^{2}}{2},
\end{equation}%
where%
\begin{equation}
w\approx \frac{\left\vert p_{1}-p_{2}\right\vert }{\sqrt{g}}\rightarrow 0.
\end{equation}

Also, 
\begin{equation}
V_{1}\rightarrow 0,V_{2}\rightarrow 0,
\end{equation}%
\begin{equation}
\cos \psi \rightarrow sign(p_{1}p_{2})\text{,}
\end{equation}%
If a particle entered the interior of\textit{\ }the horizon from its
exterior, $p>0$. If it entered from the left (mirror) region, $p<0$. Thus in
a physically relevant case when both particles came from infinity, $\psi
\rightarrow 0$.

\subsection{ $L_{1}=0$, $L_{2}=L\neq 0$}

\begin{equation}
\gamma _{12}\approx \left\vert \frac{L}{T}\right\vert \rightarrow \infty ,
\end{equation}%
\begin{equation}
w^{2}\approx 1-\frac{T^{2}}{L^{2}}\rightarrow 1\text{,}
\end{equation}%
\begin{equation}
V_{1}\rightarrow 0\text{, }V_{2}\rightarrow 1\text{,}
\end{equation}%
\begin{equation}
\cos \psi \rightarrow 0\text{.}
\end{equation}

\subsection{ \thinspace $L_{1}L_{2}>0$}

\begin{equation}
\gamma _{12}\rightarrow \frac{L_{1}^{2}+L_{2}^{2}}{L_{1}L_{2}}\text{,}
\end{equation}

\begin{equation}
w\rightarrow \frac{\left\vert L_{1}^{2}-L_{2}^{2}\right\vert }{%
L_{1}^{2}+L_{2}^{2}}<1,  \label{w+}
\end{equation}%
\begin{equation}
V_{1}\rightarrow 1\text{, }V_{2}\rightarrow 1\text{,}
\end{equation}%
\begin{equation}
\cos \psi \approx 1-\frac{T^{2}}{g}\frac{(p_{1}L_{2}-p_{2}L_{1})^{2}}{%
L_{1}^{2}L_{2}^{2}}\text{.}
\end{equation}

Particles move almost parallel to each other near the singularity.

\subsection{ $L_{1}L_{2}<0$}

\begin{equation}
\gamma _{12}\approx 2\frac{\left\vert L_{1}L_{2}\right\vert }{T^{2}}%
\rightarrow \infty ,
\end{equation}%
\begin{equation}
w\approx 1-\frac{T^{4}}{4L_{1}^{2}L_{2}^{2}}\rightarrow 1,
\end{equation}%
\begin{equation}
V_{1}\rightarrow 1\text{, }V_{2}\rightarrow 1\text{,}
\end{equation}%
\begin{equation}
\cos \psi \rightarrow -1\text{.}
\end{equation}

This means that head-on collision occurs, $\psi \rightarrow \pi $.

Now, we can summarize the results of the present section in Table 1.

\begin{tabular}{|l|l|l|l|l|}
\hline
& $L_{1}$ & $L_{2}$ & $w$ & $\psi $ \\ \hline
A & $0$ & $0$ & $0$ & $0$ \\ \hline
B & $0$ & $\neq 0$ & $1$ & $\frac{\pi }{2}$ \\ \hline
C & $\neq 0$ & $\neq 0$, \thinspace $L_{2}$ parallel to $L_{1}$ & separated
from $1$ & $0$ \\ \hline
D & $\neq 0$ & $\neq 0$, \thinspace $L_{2}$ antiparallel to $L_{1}$ & $1$ & $%
\pi $ \\ \hline
\end{tabular}

Table 1. Types of particles collisions near the singularity

\section{Particles move within different planes}

As is well-known, in the case of conserved angular momentum a particle moves
within a plane. According to above consideration, we can choose this plane
to be $\ \theta =\frac{\pi }{2}$ for, say, particle 1. However, in general,
this is not the case for particle 2, the variable $\theta $ will be varying
in time. Will it significantly affect the results for the relative velocity
and Lorentz factor $\gamma _{12}$ near the singularity? To answer this
question, we generalize the results of the previous section. Omitting the
details of derivation, we give the corresponding formulas below. Now,%
\begin{equation}
u^{\mu }=(P\text{,}-\frac{p}{g},\frac{\sigma Q}{T^{2}},\frac{L}{T^{2}\sin
^{2}\theta })\text{,}  \label{uu}
\end{equation}%
where $\sigma =\pm 1$,%
\begin{equation}
Q=\sqrt{L_{tot}^{2}-\frac{L^{2}}{\sin ^{2}\theta }}\text{,}  \label{Q}
\end{equation}%
\begin{equation}
P=\sqrt{p^{2}+g(1+\frac{L_{tot}^{2}}{T^{2}})},  \label{PT}
\end{equation}%
it is implied that%
\begin{equation}
L_{tot}\geq \frac{\left\vert L\right\vert }{\sin \theta }\text{.}
\label{tot1}
\end{equation}

Here, the integral of motion $L_{tot}$ has the meaning of the total angular
momentum of a particle, while $L$ is its component corresponding to a
variable $\phi $. Then, for $V^{(a)}=(V^{(1)},$ $V^{(2)},$ $V^{(3)})$ one
finds%
\begin{equation}
V^{((a)}=(-\frac{p}{P}\text{, }\frac{\sigma Q\sqrt{g}}{\left\vert
T\right\vert P}\text{, }\frac{L\sqrt{g}}{\left\vert T\right\vert P\sin
\theta })\text{,}  \label{va}
\end{equation}%
eq. (\ref{ga1}) is still valid but now with (\ref{PT}). Obviously,%
\begin{equation}
V_{\perp }=\sqrt{\left( V^{(2)}\right) ^{2}+\left( V^{(3)}\right) ^{2}}=%
\frac{\sqrt{g}}{\left\vert T\right\vert P}L_{tot}\text{.}
\end{equation}

We assume that for particle 1 $\theta =\frac{\pi }{2}$, $Q_{1}=0,$ $%
L_{1tot}=\left\vert L_{1}\right\vert $. Then, in the point of collision both
particles have the same coordinates, so $\theta =\frac{\pi }{2}$ for
particle 2 as well. It is convenient to introduce an angle $\alpha $ for
particle 2. so that $L_{2}=L_{tot}\cos \alpha $, where $\cos \alpha $ can
have any sign. Then, in the point of collision we have for particle 2%
\begin{equation}
u^{\mu }=(P\text{,}-\frac{p}{g},\frac{L_{2tot}\sin \alpha }{T^{2}},\frac{%
L_{2tot}\cos \alpha }{T^{2}})\text{.}
\end{equation}

Eqs. (\ref{ga12}), (\ref{gacos}) \ are also valid but in $P$ the quantity $%
L_{tot}$ appears instead of $L$.

Now,%
\begin{equation}
\cos \psi =\frac{1}{\sqrt{p_{1}^{2}+g\frac{L_{1}^{2}}{T^{2}}}}\frac{1}{\sqrt{%
p_{2}^{2}+g\frac{L_{2tot}^{2}}{T^{2}}}}(p_{1}p_{2}+\frac{L_{1}L_{2}g}{T^{2}})%
\text{,}
\end{equation}%
\begin{equation}
\gamma _{12}=\frac{P_{1}P_{2}-p_{1}p_{2}}{g}-\frac{L_{1}L_{2}}{T^{2}}\text{.}
\end{equation}

When a singularity is approached, $V^{(1)}\rightarrow 0$ as before, while $%
V_{\perp }\rightarrow 1$, so $V\rightarrow 1$ as well. Let us denote the
cases A-D depending on the $L_{1}$, $L_{2}$ in the manner similar to that in
the former section. Then, one can see that cases A and B coincide with those
from Table 1. Indeed, if one of angular momenta is zero, one can choose the
equatorial plane for another particle to be $\theta =\frac{\pi }{2}$, so
nothing new happens. Obviously, case D is similar to that from Table 1. It
remains to check what happens in case C. Then,

\begin{equation}
\cos \psi \rightarrow \frac{L_{2}}{L_{2tot}}=\cos \alpha \text{,}
\end{equation}%
$\psi =\alpha $. Taking into account that $V_{1}\rightarrow 1$ and $%
V_{2}\rightarrow 1$, we see that according to (\ref{gacos}), in case C a new
possibility arises : 
\begin{equation}
\gamma _{12}\approx \frac{\left\vert L_{1}\right\vert (L_{2tot}-L_{2})}{T^{2}%
},
\end{equation}%
so $\gamma _{12}\rightarrow \infty $ in spite of $L_{1}L_{2}>0$. Such a
possibility was absent when both particles had been moving within the same
plane (see Table 1 above). The similar phenomenon for massless particles was
discussed in \cite{h}.

\section{Motion under the action of force}

Let now some force act on a particle. Then, the equations of motion formally
retain their form but the quantities $p$ and $L$ cease to be integrals of
motion and become the functions of time. If there is an acceleration $a^{\mu
}$, one finds its tetrad components using (\ref{0}) - (\ref{3}) that
(assuming $\theta =\frac{\pi }{2}$)

\begin{equation}
a^{(3)}=\left\vert T\right\vert a^{\phi }=\frac{a_{\phi }}{\left\vert
T\right\vert },
\end{equation}%
\begin{equation}
a^{(y)}=\sqrt{g}a^{y}=\frac{a_{y}}{\sqrt{g}},
\end{equation}%
\begin{equation}
a^{(\hat{t})}=-\frac{a^{T}}{\sqrt{g}}=a_{T}\sqrt{g}.
\end{equation}%
If $\xi ^{\mu }$ is the Killing vector, it is easy to notice that 
\begin{equation}
\frac{d}{d\tau }(\xi ^{\mu }u_{\mu })=\xi ^{\mu }a_{\mu }\text{.}
\end{equation}%
Then,%
\begin{equation}
\frac{dp}{d\tau }=a_{y}=\sqrt{g}a^{(y)}\text{,}
\end{equation}%
\begin{equation}
\frac{dL}{d\tau }=a_{\phi }=\left\vert T\right\vert a^{(3)}\text{,}
\end{equation}%
where we used the same definitions $p=-u_{y}$ and $L=u_{\phi }$ as for free
particles. Now,%
\begin{equation}
\frac{p^{2}}{g}-\frac{\left( u^{T}\right) ^{2}}{g}+\frac{L^{2}}{T^{2}}=-1,
\end{equation}%
where%
\begin{equation}
u^{T}=\frac{dT}{d\tau }=\sqrt{p^{2}+g(1+\frac{L^{2}}{T^{2}})}=P\text{.}
\end{equation}%
It follows from equations of motion that%
\begin{equation}
\frac{dp}{dT}=-\frac{dp}{dr}=\frac{\sqrt{g}a^{(y)}}{\sqrt{p^{2}+g(1+\frac{%
L^{2}}{T^{2}})}},  \label{dp}
\end{equation}%
\begin{equation}
\frac{dL}{dT}=-\frac{dL}{dr}=\frac{\left\vert T\right\vert a^{(3)}}{\sqrt{%
p^{2}+g(1+\frac{L^{2}}{T^{2}})}}.  \label{dL}
\end{equation}%
It is clear from (\ref{dp}), (\ref{dL}) that $p$ and $L$ remain finite, if $%
a^{(y)\text{ }}$and $a^{(3)\text{ }}$are finite.

This has important consequences for the properties of velocities. In
particular, in the tetrad (\ref{0})$\ $- (\ref{3}) $V^{(1)}\rightarrow 0$
and $V^{(3)}\rightarrow \pm 1$, if $L\neq 0$ ($V^{(3)}=0$ for $L=0$). These
conclusions are valid for any finite $E$, $L$\cite{rad}, so they apply to
the case under discussion as well. Therefore, we come to an important
conclusion: the presence of finite force does not abolish the effect of high
energy collision. This statement is the counterpart of similar results for
the BSW effect near the horizon \cite{tzf}.

\section{When particle velocity can approach the speed of light}

It follows from the above consideration that eq. (\ref{V}) indeed retains
its validity, if the constants of motion $p$ and $L$ are replaced by their
momentary values $p(T)$ and $L(T).$ In turn, this has an important
consequence. For a finite acceleration, the velocity can reach the limiting
value $V=1$ only in two cases: when approaching the horizon and/or
singularity. In the first case, the right hand side of (\ref{V}) diverges
due to the first term where $g\rightarrow 0$. In the second one it does so
due to the second term where $T\rightarrow 0$.

All these conclusions are obtained with the assumption that $a_{(i)}$ are
finite and hence $p$ and $L$ are finite as well. If we relax the requirement
of finiteness of $a_{(i)}$, an additional possibility opens that $%
V\rightarrow 1$ due to unbounded acceleration and, correspondingly,
unbounded $p$ and $L$. Thus there are three possibilities for getting $%
V\rightarrow 1$: (i) horizon, (ii) singularity, (iii) infinite acceleration.

This result is valid for the velocities with respect to the Lema\^{\i}tre
frame as well. Moreover, it is valid with respect to a general radially free
falling system formed by particles with the specific energy $e_{0}$. It is
known that in such a general case the radial component of velocity of a
particle with specific energy $e$ is given by (see \cite{rad}) 
\begin{equation}
V^{(1)}=\frac{P_{0}e-Pe_{0}}{e_{0}e-PP_{0}}  \label{v1e}
\end{equation}

and the angular component is

\begin{equation}
V^{(3)}=\frac{\mathcal{L}f}{r(e_{0}e-PP_{0})}\text{,}  \label{v3e}
\end{equation}

where 
\begin{equation}
P_{0}=\sqrt{e_{0}^{2}-f}.  \label{P}
\end{equation}

Formulae of the present paper for the components of velocity of an
individual particle can be thought of as a particular case $e_{0}=0$
(corresponding to a resting observer with $y=const$ under the horizon) of
these general formulae. Assuming that $e$ is finite, we see that $V^{(3)}<1$
as it should be and $V^{(3)}\rightarrow 1$ at singularity. As for the radial
component, substituting $P$ and $P_{0}$ into the condition $V^{(1)}=1$ and
considering finite $e$ we get, after a simple algebra, that $f=0$. This
means that $V^{(1)}$ can take the value $1$ at the horizon only, provided $e$
is finite. It is worth noting that this result is valid for any spherically
symmetric static space-times since we do not specify the function $f$. The
coordinate system $e_{0}=0$ considered here becomes singular at the horizon
itself. The discussion contained in Sec. VI of \cite{rad} explains, why $%
V^{(1)}=1$ for other systems, regular at a horizon, is safe from physical
point of view (note, that unlike a singularity we have equality, not a limit
here!).

\section{Another types of space-like singularity: finite $g$}

Up to now, we considered the singularity of the Schwarzschild type. This
implied that in the limit when $T=-r\rightarrow 0$, the scale factor $%
g\rightarrow \infty $. In this section, we consider what happens if $g$
remains finite there. In particular, this case is encounted in so-called
T-models \cite{rub}. Note, that we still assume a metric without inner
horizons. Now, in (\ref{ga12}) the only potential source of divergences is
the term with $T$ in the denominator. Analyzing this expression and (\ref{v1}%
, (\ref{v3}), we arrive at the following conclusions.

Consider motion of two particles within the same plane. If a particle has $%
L=0$, $\left\vert V^{(1)}\right\vert \rightarrow \frac{\left\vert
p\right\vert }{\sqrt{p^{2}+g(0)}}<1$, $V^{(3)}\equiv 0$. This differs from
the singular case where $V^{(1)}$ tends to zero. However, if $L\neq 0$, $%
\left\vert V^{(1)}\right\vert \rightarrow 0$, $V^{(3)}\rightarrow \pm 1$ as
in the singular case. Correspondingly, if $L_{1}=0=L_{2}$, $\gamma _{12}$ is
finite (though does not equal to $1$ in the case when particles have
different values of $p$). If $L_{1}=0$, $L_{2}\neq 0$, $\gamma
_{12}\rightarrow \infty $. If $L_{1}L_{2}<0$, $\gamma _{12}\rightarrow
\infty $ as well. If $L_{1}L_{2}>0$, $\gamma _{12}$ is finite.
Straighforward calculation shows that the formula (32) for a relative
velocity changes quantitatively: 
\begin{equation}
w=\frac{|L_{2}^{2}(1-p_{1}/g(0))-L_{1}^{2}(1+p_{2}/g(0))|}{%
L_{2}^{2}(1+p_{1}/g(0)+L_{1}^{2}(1+p_{2}/g(0))}.  \label{g0}
\end{equation}%
In the limit $g(0)\rightarrow \infty $ eq. (\ref{g0}) turns into eq. (\ref%
{w+}) as it should be. Comparing with Table 1, we see that there is no
qualitative difference between the Schwarzschild case and the one under
discussion.

\section{Generalizing Lema\^{\i}tre reference frame}

In the above consideration, we mentioned the Lema\^{\i}tre frame that is
used sometimes for the description of the black hole interior. Its standard
use implies that reference particles compose the set of observes free
falling from infinity from the state of rest, so their specific energy $%
e_{0}=1$. Meanwhile, this frame can be generalized if the role of reference
particles is played by observers with $e_{0}\neq 0$. This is done in \cite%
{rad} where even cases with $e_{0}<0$ and $e_{0}=0$ were included in a
general scheme. This leads to formulae like (\ref{v1e}) and (\ref{v3}). Now,
we make the next step and build the generalized Lema\^{\i}tre frame based on
particles with nonzero specific angular momentum $L_{0}$. For brevity, we
call it $L$-Lema\^{\i}tre frame. It is relevant, in particular, under the
horizon where, as we saw, radial particle motion is unstable, so if we want
to know how dynamics looks like from the viewpoint of a falling particle, it
is natural to use the $L$-Lema\^{\i}tre frame. Below, we list main formulas
for it. In principle, in order to describe mutual motion of two freely
falling particle, we can use a standard frame and perform the Lorentz
rotation, as we did when we got the singularity asymptotic for mutual
velocity. Considering the $L$-Lema\^{\i}tre frame is an alternative and more
direct way to get the results.

As consideration runs along the same lines as above and in \cite{rad}, we
give the main formulas briefly.

We choose the reference particle to move in the plane $\theta =\frac{\pi }{2}
$ and restrict ourselves by the region under the horizon. We attach to it a
zeroth vector of a tetrad and choose the other three ortogonal to it. In
coordinates $(T,y,\theta $, $\phi )$

\begin{equation}
u_{0}^{\mu }=h_{(0)}^{\mu }=(P_{0}\text{,}-\frac{p_{0}}{g},0,\frac{L_{0}}{%
T^{2}})\text{,}
\end{equation}%
\begin{equation}
h_{(0)\mu }=(-\frac{P_{0}}{g},-p_{0},0,L_{0}).  \label{h0}
\end{equation}%
Here, the parameters of a reference particle have the subscript 0,%
\begin{equation}
P_{0}=\sqrt{p_{0}^{2}+g(1+\frac{L_{0}^{2}}{r^{2}}})\text{.}
\end{equation}%
Also,%
\begin{equation}
h_{(1)\mu }=\frac{1}{\sqrt{1+\frac{L_{0}^{2}}{r^{2}}}}(\frac{p_{0}}{g}%
,P_{0},0,0),
\end{equation}%
\begin{equation}
h_{(2)\mu }=\left\vert T\right\vert (0,0,1,0),
\end{equation}%
\begin{equation}
h_{(3)\mu }=(-\frac{P_{0}L_{0}}{g\left\vert T\right\vert \sqrt{1+\frac{%
L_{0}^{2}}{r^{2}}}},-\frac{p_{0}L_{0}}{\left\vert T\right\vert \sqrt{1+\frac{%
L_{0}^{2}}{r^{2}}}},0,r\sqrt{1+\frac{L_{0}^{2}}{T^{2}}})\text{.}
\end{equation}%
Then, using the general definition \cite{72}%
\begin{equation}
V^{(i)}=V_{(i)}=-\frac{h_{(i)\mu }u^{\mu }}{h_{(0)\mu }u^{\mu }},
\end{equation}%
we have for a freely falling particle%
\begin{equation}
V^{(1)}=\frac{1}{\gamma g\sqrt{1+\frac{L_{0}^{2}}{T^{2}}}}(Pp_{0}-P_{0}p),
\end{equation}%
\begin{equation}
V^{(2)}=\frac{\sigma Q}{\gamma \left\vert T\right\vert }
\end{equation}%
\begin{equation}
V^{(3)}=\frac{1}{\gamma \left\vert T\right\vert \sqrt{1+\frac{L_{0}^{2}}{%
T^{2}}}}(\frac{L}{\gamma }-L_{0}).
\end{equation}%
where $\gamma $ is the individual Lorentz gamma factor of a particle%
\begin{equation}
\gamma =-u_{(0)\mu }u^{\mu }\text{.}
\end{equation}

It follows from (\ref{u}), (\ref{uu}), (\ref{h0}) that 
\begin{equation}
\gamma =\frac{P_{0}P-pp_{0}}{g}-\frac{LL_{0}}{T^{2}}\text{.}
\end{equation}
The quantity $Q$ is given by eq. (\ref{Q}) and contains $L_{tot}$.

These formulae give \textit{the most general} expressions for 3-velocity
components with respect to an arbitrary freely falling frame. At a
singularity they reduce to the results obtained above.

In is worth to note that in the case of $L_0=L_1$ and two equatorial planes
coinciding, the angular component of velocity $V^{(3)}$ vanishes only in a
singularity, being non-zero everywhere else (see Fig.1), so the Eq.(32) is
the limiting equation even if two angular moments are the same. 
\begin{figure}[tbp]
\includegraphics[scale=0.8]{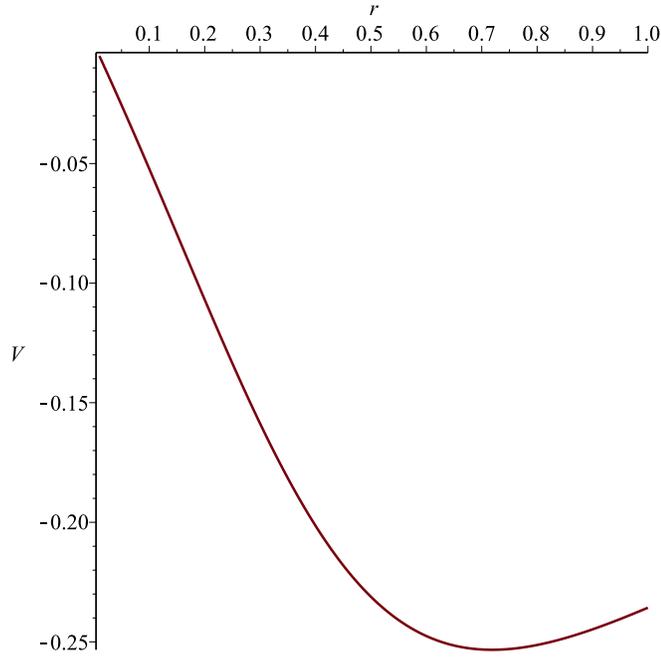}
\caption{The angular component $V^{(3)}$ of velocity of a particle with $%
L=m^{2}$ and $e=2$ with respect to the frame with $L_0=m^{2}$ and $e_0=1$.
The unit for $V^{(3)}$ is $c$.}
\label{Fig1}
\end{figure}

\section{Motion with respect to a frame versus motion of nearby particles}

Apart from the behavior of velocity with respect to a fixed frame, another
interesting question is mutual movement of nearby points. The properties of
such a motion can be very different from the motion with respect to a fixed
frame, even if one of the points considered is at rest with respect to the
coordinate system in question. The reason is that the velocity with respect
to a frame is a local entity, while a distance between two points is a
space-like variable. In the present section we consider both local and
non-local velocities. The goal of the section is mostly pedagogical since it
contains a large part of known issures, though reinterpreted in our
notations.

First, consider the radial motion. The velocity along the leg of a
hypercylinder in metric (\ref{met}) is known to decrease and vanish in a
singularity \cite{flows}, \cite{rad}. As for the distance to a nearby point,
it increases as it can be clearly seen from the form of the metrics (\ref{1}%
), (\ref{met}). At the singularity the proper distance diverges. This means
that a velocity of nearby (but anyway located in a different point!)
particle should diverge if we consider "velocity" as a rate of change of a
proper distance.

A reasonable way to define a velocity of a distant object is to consider the
derivative of a proper distance with respect to a \textit{proper} time of an
observer. The metric (\ref{met}) is not a synchronous one, so we introduce a
proper time of an observer being at rest from the viewpoint of the metric (%
\ref{met}) as 
\begin{equation}
d\tilde{t}=dT/\sqrt{g}.  \label{time}
\end{equation}%
With this definition we have for the velocity in question 
\begin{equation}
\frac{dl}{d\tilde{t}}=\frac{d(\sqrt{g}y)}{d\tilde{t}}=\frac{d\sqrt{g}}{d%
\tilde{t}}y+\sqrt{g}\frac{dy}{d\tilde{t}}=\frac{1}{2}\frac{dg}{dT}y+V^{(1)},
\label{dl}
\end{equation}%
since $\sqrt{g}dy/d\tilde{t}=\sqrt{g}(dy/dT)(dT/d\tilde{t})=-p/(P)=V^{(1)}$,
where eqs. (\ref{u}) and (\ref{v1}) were used. So that, we got a relation
resembling the known one for cosmology -- the overall radial velocity is
decomposed into a non-local part, which can be rewritten as $(1/2\sqrt{g}%
)(dg/dT)l$ being proportional to the proper distance $l=\sqrt{g}y$ to the
particle (an analog of Hubble velocity in cosmology), and a local "peculiar
velocity" $V^{(1)}$. The later term tends to zero in a singularity while the
first term diverges (being a non-local entity, it is not bound by the speed
of light). The picture is similar to the Big Rip cosmological singularity,
apart from the fact that the Big Rip is isotropic.

In popular books, when describing influence of tidal forces to an unhappy
observer falling freely into a black hole, authors usually illustrate the
text by emotional pictures of an observer "spaghettified" in the direction
toward a singularity. This has no sense in coordinates like $T$,$y$ since
(i) they are homogeneous inside a horizon, and (ii) a singularity, being
space-like and in absolute future for an observer, is not present in any of
an observer's $T=const$ slices. By itself, "spaghettization" does occur but
has another meaning: if we make a series of snapshots of cross-sections $%
T=const$ for different $T$, the object extends more and more when $T$ grows
approaching $T=0$.

Another picture arises in the Lema\^{\i}tre coordinates. Singularity is
present in the sections of constant Lema\^{\i}tre time, so the direction
towards a singularity makes sense\textbf{. }Since for two radially separated
particles the singularity occurs at different moments of the proper time $%
\tau $ along the trajectory, the separation between two points reaches its
finite maximum when the "inner" particle hits a singularity (in this context
the word "hits" is conditional since the singularity is space-like, we use
it for brevity only).

It is easy to estimate this maximum for a pure radial motion. For the
Schwarzschild black hole this can be a student seminar exercise. Suppose two
particles, being at rest with respect to the Lema\^{\i}tre system are
separated by some distance. Let a particle move with $E=m$, so it would
start its motion from the rest at infinity. Then, it is known (see, e.g. eq.
2.3.12 of \cite{fn}) that if a particle moves from some $r$ to $r_{f}<r$,
the proper time $\tau$ is equal to 
\begin{equation}
\tau (r,r_{f})=(2/3)r_{g}^{-1/2}(r^{3/2}-r_{f}^{3/2}).  \label{tauf}
\end{equation}%
In particular, the proper time between a given position $r$ and the
singularity $r=0$ is obtained from (\ref{tauf}) if we put there $r=0$, so%
\begin{equation}
\tau (r,0)=(2/3)r_{g}^{-1/2}r^{3/2}.  \label{tay}
\end{equation}

It is also worth mentioning that for such a particle the Lema\^{\i}tre time
coincides with the proper one (see, e.g. eq. 14 of \cite{flows}). Also, the
proper distance in this case is equal to the difference of the coordinate
values of $r$.

Let we have two such particles initially separated by the coordinate
distance $l$. We want to find the location $r_{f}$ of the "outer" particle
initially located at $r+l$, at the moment $\tau $ when the "inner" particle
hits the singularity. Equating $\tau (r+l,r_{f})=\tau (r,0$), assuming small 
$l$ and expanding the right hand side with respect to $l/r$ we get 
\begin{equation}
r_{f}=[(3/2)l]^{2/3}r^{1/3},
\end{equation}%
which gives for the ratio 
\begin{equation}
r_{f}/l=(3/2)^{2/3}(r/l)^{1/3}.
\end{equation}%
Thus small absolute displacements remain small. However, relative
displacement may be arbitrary large.

As a trilling example we can consider the following situation: suppose that
different parts of human body ($l\sim 1m$) start to move geodesically after 
\textit{tidal} acceleration $g_{t}$ exceeds the free fall acceleration at
the surface of Earth ($g_{E}\sim 10m/s^{2}\sim 10^{-16}m^{-1}$ in natural
units $c=1$). For small $l$, $g_{t}\approx r_{g}l/r^{3}$ (see, e.g. page
B-20 of \cite{TW}). Thus free fall begins at $r=(r_{g}l/g_{E})^{1/3}$. Using
these data we can estimate 
\begin{equation}
r_{f}/l\sim 60r_{g}^{1/9},  \label{sp}
\end{equation}%
where $r_{g}$ is expressed in meters. This indeed indicates "spagettization"
- the size in $r$-direction enlarges from about 100 times for a stellar mass
black hole to about 1000 times for a supermassive ($10^{9}$ solar masses)
one (note, that the dependence upon black hole mass is rather weak).

We can add also that in the Lema\^{\i}tre coordinates the coordinate
velocity itself admit a decomposition 
\begin{equation}
dr/d\tilde \tau=-v+V_{(r)}
\end{equation}%
where $\tilde \tau$ is the Lema\^{\i}tre time, $v=\sqrt{1-f}$ and $V_{r}$ is
a local velocity \textit{with respect to the Lema\^{\i}tre frame} (see the
equation (54) in\cite{flows}). It should not to be confused with $V^{(1)}$
in (\ref{dl}) where the frame with $p=0$ was used. Nevertheless, the
asymptotic of the terms in the right hand side of this equation at a
singularity are the same: $v \to \infty$ and $V_{(r)} \to 0$ (see
corresponding formulae in \cite{flows}). On the other hand, in a direct
analog of (\ref{dl}) the non-local flow velocity (the first term in the
decomposition) is equal to difference of the values of $v$ at the positions
of the particle and the observer, and this difference is, in general, not
factorizable, which means that it is no longer proportional to the proper
distance between the particle and the observer.

For the motion in the angular direction the situation is quite opposite.
Suppose we have a particle with a zero angular momentum, so it falls along $%
\phi =0$ line, and a nearby particle does so with some small but non-zero $L$%
. We know that $V^{(3)}$ of the second particle tends to $1$ when the
singularity is approached. Does this mean that the proper distance between
these two particles increase rapidly? The answer is "no" as the direct
dependence $\phi (r)$ in the Schwarzschild metric shows (Fig.2). In this
picture we plot $V^{(3)}$ of a particle with $L=m^{2}$ inside a horizon. It
tends to $1$ near a singularity. In the same plot we show the distance from
the line $\phi =0$ to this particle (we assume that this particle crosses
the line $\phi =0$ at the horizon) which is equal to $r\phi =-T\phi $. This
distance first increases due to non-zero $L$ (as it would be in a flat space
also), then it starts to decrease despite growing velocity $V^{(3)}$. The
contraction in the angular direction overcomes, and the distance in the $%
\phi $ direction appears to be always smaller than it would be without
gravity.

This picture is qualitatively the same in the Lema\^{\i}tre coordinates as
well. The only difference is that $V^{(3)}$ in static coordinate always
vanishes at a horizon (this is a counterpart of the statement that radial
velocity is always $1$ at a horizon), while the analog of this value with
respect to the Lema\^{\i}tre system can take any value from $0$ to $1$.

As for the angle $\phi $ itself, it reaches a finite value at singularity.
This value grows with growing $L$, tending to $\pi $ for $L\rightarrow
\infty $ (see eq.21 of \cite{rad19})\textbf{.}

\begin{figure}[tbp]
\includegraphics[scale=0.8]{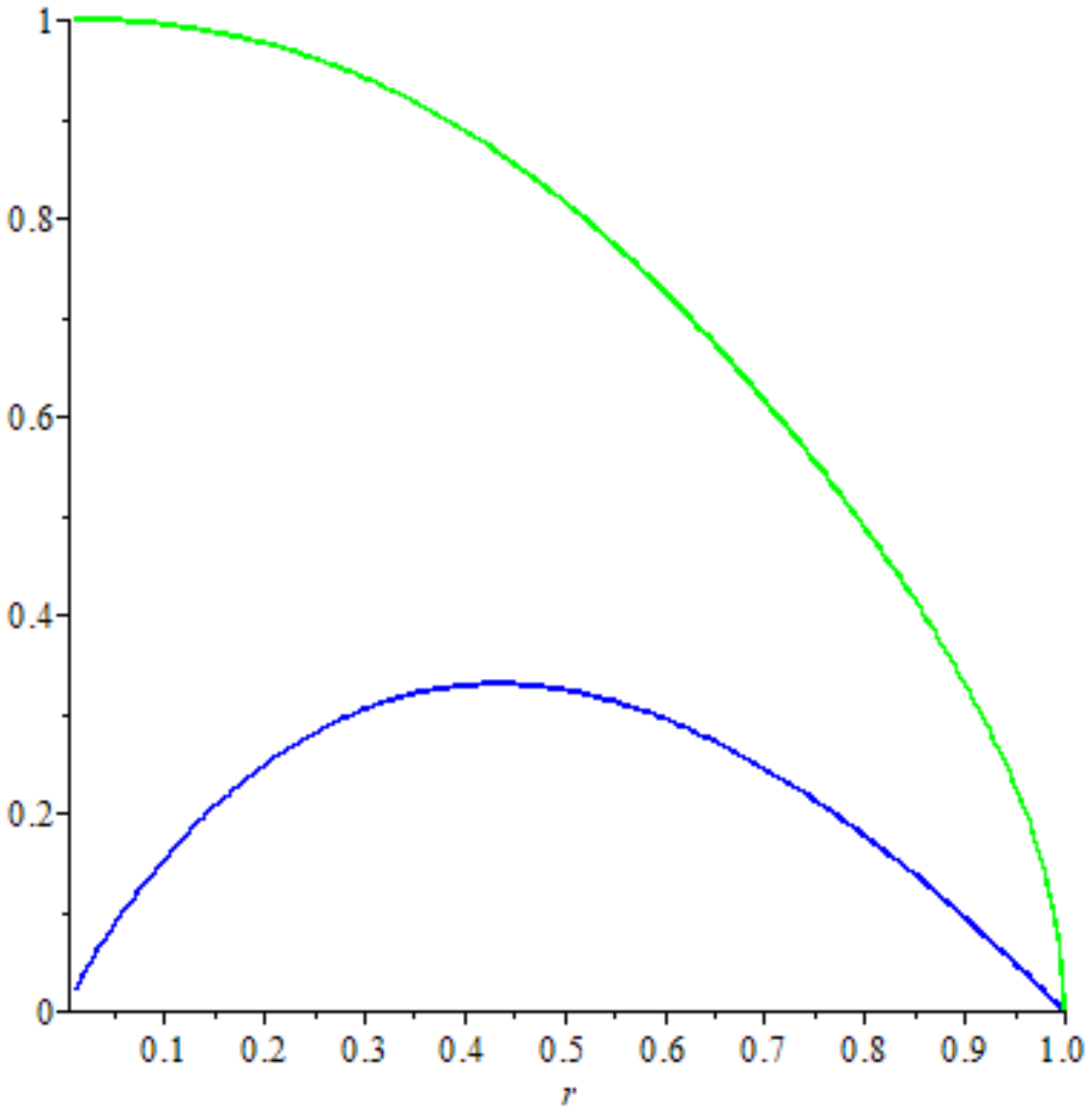}
\caption{The angular component $V^{(3)}$ of velocity of a particle with $%
L=m^{2}$ inside a horizon (green) and distance to the particle in angular
direction from the radius $\protect\phi =0$ crossed by this particle at a
horizon (blue). The unit for $V^{(3)}$ is $c$, the unit for the distance is $%
r_{g}$.}
\label{Fig2}
\end{figure}

It is worthwhile to mention that static coordinates admit a "Hubble-like"
decomposition for angular motion as well. Indeed, if the first particle has $%
\phi =0$, then using Eq.(\ref{time}), (\ref{u}) and (\ref{v3}), we have 
\begin{equation}
\frac{dl}{d\tilde{t}}=-\frac{d(T\phi )}{d\tilde{t}}=-\sqrt{g}\phi +\frac{L%
\sqrt{g}}{TP}=-\sqrt{g}\phi +V^{(3)},
\end{equation}%
so an analog of the "Hubble flow" is directed inwards, its velocity is
proportional to $\phi $ of the second particle, and if $g(0)$ is singular,
the corresponding velocity diverges. This means that it dominates near a
singularity since $V^{(3)}$, being a physical velocity, is bounded by the
speed of light while the first term does not.

This decomposition takes place only for the "static" coordinates
(corresponding to the observer with $p=0$, $y=const$), in the Lema\^{\i}tre
coordinates the spatial sections are flat, so the distance in angular
direction is $l=r\sin {\phi }$, and the above property is lost since the
angular component of the peculiar velocity in the Lema\^{\i}tre frame does
not depend on the value of $\phi$ (see \cite{flows}) while the second term
in the decomposition of $dl/d\tilde \tau$ does.

\section{Discussion and Conclusions}

In this paper we have considered dynamical phenomena in the vicinity of the
singularity of the Schwarzschild-like space-time. This concerns not only the
Schwarzschild metric but any singularity of the same type when $r\rightarrow
0$ and $g\rightarrow \infty $. In this case one can regard horizon's
interior as an anisotropic, dynamical $V^{4}$ space-time with a
hypercylinder $V^{3}=R^{1}\times S^{2}$ space-like sections. There is a
longitudinal, $R^{1}$-expansion and transversal, $S^{2}$-contraction. Due to
extremely violent $R^{1}$-expansion in its final stage one could expect the
asymptotic state of mutual rest of all the particles moving along $y$%
-direction\ (see e.g \ \cite{rad19}). This picture has been completed by a
Doppler's blueshift, for the case of transverse component of trajectories: a
light-like, non-zero angular momentum geodesics have been recorded
blueshifted \cite{gr20}. We have verified here the kinematics of the test
particles following time-like, non-zero angular momentum trajectories of
both geodesic and non-geodesic character. If a test particle moves along an
arbitrary non-zero angular momentum trajectory, then its speed as measured
by resting observers, those with constant spatial coordinates, approaches
that of light, $w\rightarrow 1$ as $T\rightarrow 0.$ Previously, it was
found that this is valid for geodesic trajectories \cite{flows}, \cite{rad}.
Now, we showed that this is valid for an arbitrary finite force. Moreover,
the presence of a finite force is compatible with high energy collisions
near the singularity.

If, instead of one particle, we take the two particles following non-zero
angular momenta trajectories, their relative velocity $w\rightarrow 1$ with
only one exceptional case. It occurs if both particles move in the same
plane and have parallel angular momenta; then the value\textbf{\ }of\textbf{%
\ }their relative speed $w$ is smaller than that of light, $w<1$. This also
happens if both particles have zero angular momenta. Otherwise, non-zero
angular momentum of a test particle is a necessary and sufficient condition
for $w\rightarrow 1.$

It should be pointed out that there exists the reason, common for both the
indefinite blueshift for the class of non-zero angular momentum light-like
geodesics and indefinite tendency of the relative speed of the particles
following their non-zero angular momentum trajectories to the speed of light
when approaching the ultimate singularity $T\rightarrow 0$ of Schwarzschild
BH's interior. This effect is caused by a contraction in the course of
highly anisotropic dynamics of space-time. Indeed, when approaching $%
T\rightarrow 0,$ \cite{rub} the hypercylinder is critically contracting,
i.e. the radius $\left\vert T\right\vert $ of the two-sphere, diminishes to
the zero value, $T\rightarrow 0.$ This critical contraction carries all of
the objects, massive and massless, in such a way that the light recorded by
a resting or moving along $y$-axis observer turns out to be indefinitely
blueshifted and the speed of a test particle as measured by resting or
moving along $y$ axis observer tends indefinitely to the speed of light, $%
w\rightarrow 1$. When two colliding massive particles follow their non-zero
angular momenta trajectories, then in general they experience head-on
collision and their relative speed approaches that of light. (For motion
within the same plane and the same directions of the angular momenta the
effect is moderate: the relative speed of the colliding particles is found
to be smaller than the speed of light, $w<1.$\ This is an analogy of the
finding in \cite{gr20} where for motion within the same plane and the same
directions of the angular momenta of the observer and the light a finite
blueshift is found).

The result $w\rightarrow 1$ may be regarded as a center of mass energy
collision tending to infinity. This interpretation provides a particular
perspective. All of the variety of the BSW effect, unbounded energy
collisions in the vicinity of the black hole horizons, outer or inner, have
lead to the conclusion about arbitrary large limit which, however, is not
reached in any particular collision, so an infinite limit cannot be
realized. This is called a principle of kinematic censorship \cite{cens}.
Meanwhile, in the case under discussion this principle is violated when $%
T\rightarrow 0$ ($r\rightarrow 0$). This is probably quite natural since in
the singularity itself all known laws of physics can be violated and
geometry as such ceases to exist.

Since exact vanishing of angular momentum and exact coinciding of planes of
motion for two particles represent zero-measure set of initial conditions
and cannot be exactly satisfied in any realistic physical situation, we can
conclude that tending $w$\ to $1$ at a singularity is unavoidable.
Correspondingly, indefinite growth of $E_{c.m.}$ is general feature for
particle collisions near the singularity.

It is also shown for spherically symmetric space-times of a quite general
form that a particle velocity can approach the speed of light only in three
cases: (i) on the horizon, (ii) in the singularity, (iii) when a proper
acceleration diverges.

We also considered another type of singuarity when $g(0)$ is finite. This
applies, for example to T-models of sphere that represent a special type of
the Kantowski-Sachs metric. This includes the solution of for the collapse
of dust complimentary to the LTB models \cite{rub}. The comparison of
particle dynamics near both types of singulrity is carried out, it showed
the possibility of high energy collisions in both cases.

We also generalized the Lema\^{\i}tre type of frame which is built from
particles with nonzero angular momenta. It can be useful for description of
particle flows under the horizon.

The most important and intriguing issue that arises from our finding is the
question about stability or instability of the vacuum singularity if
backreaction of matter is taken into account. This remains an open question.

\section*{Acknowledgement}

O. Z. thanks H. V. Ovcharenko for useful discussion and Wroclaw University
of Science and Technology where this work started. A.T. and O. Z. thank for
hospitality the Center for Theoretical Physics in British University, Cairo,
where this work was finished. 
The work of AT is supported by the Program of Competitive Growth of Kazan
Federal University and by the Interdisciplinary Scientific and Educational
School of Moscow University in Fundamental and Applied Space Research.






\end{document}